\documentclass[runningheads]{svmult}
\usepackage{makeidx}   
\usepackage{graphicx}  
\usepackage{subeqnar}  
\usepackage{multicol}  
\usepackage{cropmark} 
\usepackage{physprbb}  
\newcommand{\be}{\begin{equation}}
\newcommand{\ee}{\end{equation}}
\newcommand{\bea}{\begin{eqnarray}}
\newcommand{\eea}{\end{eqnarray}}

\newcommand{\AmS}{{\protect\the\textfont2
  A\kern-.1667em\lower.5ex\h\citebox{M}\kern-.125emS}}
\newcommand{\lsim}{\mathrel{\mathop{\kern 0pt \rlap
  {\raise.2ex\hbox{$<$}}}
  \lower.9ex\hbox{\kern-.190em $\sim$}}}
\newcommand{\gsim}{\mathrel{\mathop{\kern 0pt \rlap
  {\raise.2ex\hbox{$>$}}}
  \lower.9ex\hbox{\kern-.190em $\sim$}}}
\def\Journal#1#2#3#4{{#1} {#2} (#3) #4 }

\def\NCAA{{\em Il Nuovo Cimento} A}

\def\PLB{{\em Phys. Lett.} B}

\def\PRL{\em Phys. Rev. Lett.}

\def\PRD{{\em Phys. Rev.} D}

\begin{document}

\title*{Results from DAMA/NaI and perspectives for DAMA/LIBRA}
\toctitle{Results from DAMA/NaI and perspectives for DAMA/LIBRA}

%
%
\titlerunning{Results from DAMA/NaI and perspectives for DAMA/LIBRA}
%
\author{R. Bernabei\inst{1}
\and P. Belli\inst{1}
\and F. Cappella\inst{1}
\and F. Montecchia\inst{1} \thanks{also:
Universit\`a "Campus Bio-medico" di Roma, 00155, Rome, Italy}
\and F. Nozzoli\inst{1}
\and A.~Incicchitti\inst{2}
\and D. Prosperi\inst{2}
\and R. Cerulli\inst{3}
\and C.J. Dai\inst{4}
\and H.H. Kuang\inst{4}
\and J.M. Ma\inst{4}
\and Z.P. Ye\inst{4} \thanks{also:
University of Zhao Qing, Guang Dong, China}}
\authorrunning{R. Bernabei et al.}
%
%
\institute{Dipartimento di Fisica, Universit$\grave{a}$ di Roma
``Tor Vergata'' and INFN, Sezione di Roma2, I-00133 Rome, Italy
\and Dipartimento di Fisica, Universit$\grave{a}$ di
Roma ``La Sapienza'' and INFN, Sezione di Roma, I-00185 Rome, Italy
\and INFN - Laboratori Nazionali del Gran Sasso, I-67010 Assergi (Aq), Italy
\and IHEP, Chinese Academy, P.O. Box 918/3, Beijing 100039, China}

\maketitle 

\begin{abstract}
The $\simeq$ 100 kg highly radiopure NaI(Tl) set-up of the DAMA project (DAMA/NaI) 
took data over seven annual cycles up to 
July 2002 and has achieved results on various rare processes. Its main 
aim has actually been 
the exploitation of the model independent WIMP annual modulation signature.
After this conference the total exposure, collected 
during the seven annual cycles, was released.  
This cumulative exposure (107731 kg $\times$ day) has given 
a model independent evidence for the presence of a Dark Matter particle component
in the galactic halo at 6.3 $\sigma$ C.L.; this main result is summarised here.
Some of the many possible corollary model dependent quests for the candidate particle are
mentioned. At present, after about five years of new developments, a second 
generation low background set-up (DAMA/LIBRA with a mass of $\simeq$ 
250 kg NaI(Tl)) was built and is taking data since March 2003. 
New R\&D efforts toward a possible NaI(Tl) ton set-up, we 
proposed in 1996, have been funded and started in 2003.
\end{abstract}

\vspace{-14.cm}
\begin{flushright}
{\bf to appear on the Proceed. of the Conf. "Beyond the Desert 03" \hspace{0.1cm} $\,$ \\}
{\bf Rindberg Castle, Germany, June 2003 \hspace{0.1cm} $\,$ \\}
\end{flushright}
\vspace{12.cm}

\section{Introduction}

The DAMA project has been proposed by the italian group 
to INFN and firstly funded in 1990
\cite{1}; in 1992 the chinese colleagues joined the project.
DAMA is an observatory for rare processes based on the 
development and use of
various kinds of radiopure scintillators.
Several low background set-ups have
been realised; the main ones are:
i) DAMA/NaI ($\simeq$ 100 kg of
radiopure NaI(Tl)), which took data underground over seven annual cycles and was put out of
operation in July 2002; 
ii) DAMA/LXe ($\simeq$ 6.5 kg liquid Xenon), which has been upgraded various times and is 
in operation;
iii) DAMA/R\&D, which is a set-up
devoted
to measurements on prototypes/small--scale--experiments and is in operation; 
iv) the new second generation set-up 
DAMA/LIBRA ($\simeq$ 250 kg;
more radiopure NaI(Tl)) in operation since March 2003.
Moreover, in the framework of devoted R\&D for higher radiopure
detectors and PMTs, sample measurements are regularly 
carried out by means of
the low background DAMA/Ge detector, installed deep 
underground since $\gsim$ 10 years
and, in some cases, by means of Ispra facilities.
Recent results from DAMA/LXe are presented in these 
proceedings elsewhere \cite{2}, while in ref. \cite{3} 
some recent results achieved with 
DAMA/R\&D can be found. 

The DAMA/NaI set-up and its performances have been described in details
 in ref. \cite{Nim98}.
Since then some upgrading has been carried out; 
in particular, in summer 2000 the electronic chain and
data acquisition system were completely substituted as 
reported in ref. \cite{RNC}.

As mentioned, DAMA/NaI has taken data over seven annual cycles up to July 2002 \cite{RNC}. 
The first data release and publication of DAMA/NaI data 
occurred in 1996 \cite{Psd96}
for an exposure of 4123 kg $\times$ day (DAMA/NaI-0) analysed by 
using the pulse shape discrimination (PSD) 
technique in NaI(Tl) in 
order to discriminate between electromagnetic background 
and recoils\footnote{The NaI(Tl) Dark Matter experiments have 
discrimination capability as largely stated in literature by several groups 
and at present exploited as main goal
by the NAIAD experiment.
In Zeplin I the pulse shape 
discrimination technique has been employed in LXe as well (sometimes this rejection procedure
has been indicated as rejection by {\it timing}),
similarly as done and published by DAMA/LXe in 1998. 
Moreover, we remind that 
the WIMP annual modulation signature acts (as pointed out since
the 80's by \cite{Freese}) itself as a powerful background rejection
procedure and that it can be effectively exploited 
only in large mass experiments such as so far DAMA/NaI and DAMA/LIBRA.}.
Investigation of possible diurnal effects has been carried out as well \cite{Diu}. 
The main aim of the experiment was actually the 
investigation of the presence of a Dark Matter particle component in the
galactic halo by means of the model independent WIMP annual modulation signature, which has been  
 deeply investigated over seven annual cycles (107731 kg $\times$ day total exposure).

DAMA/NaI has also obtained results on the investigation of some
exotic Dark Matter candidates (such as neutral SIMPs,
neutral nuclearities and Q-balls) \cite{exotic}, has
searched for solar axions \cite{Dax} and has given results on
several other rare processes
such as e.g. 
spontaneous emission
of protons in $^{23}$Na and $^{127}$I with violation of the Pauli exclusion principle \cite{35},
nuclear level excitation of $^{23}$Na and $^{127}$I 
during charge-non-conserving
processes \cite{cnc}, electron stability
and non-paulian transitions in Iodine atoms by L-shell\cite{Ela99}.

In the following, the result obtained by DAMA/NaI in the investigation 
of the WIMP annual modulation signature 
over the seven annual cycles will be briefly summarised.
This cumulative 107731 kg $\times$ day exposure has given
a model independent evidence for the presence of a Dark Matter particle component
in the galactic halo at 6.3 $\sigma$ C.L.;
this main result and corollary interpretations in some of the possible different model 
scenarios have been discussed in experimental 
and theoretical details in ref.
\cite{RNC} and have firstly been presented after this Conference in a seminar
at Gran Sasso National Laboratory (LNGS) at end of July 2003.

The new second generation larger mass and  
higher radiopure NaI(Tl) set-up, named DAMA/LIBRA, is now in operation
and will be briefly introduced.

\section{Some generalities}

The DAMA/NaI 
experiment has been built in order to have suitable
 mass, sensitivity and control
of the running conditions to investigate the WIMP 
presence in the galactic halo
by the annual modulation signature.
We remind that the WIMP annual modulation signature
 is based on the annual modulation of
the signal rate induced by the Earth revolution
around the Sun; as a consequence, the Earth will be 
crossed by a larger WIMP flux roughly in June 
(when its rotational velocity 
is summed 
to the one of the solar system with respect to the Galaxy)
and by a smaller one roughly in December
 (when the two velocities are subtracted).
The annual modulation signature is very distinctive
since a WIMP-induced seasonal effect must simultaneously satisfy
all the following requirements: the rate must contain a component
modulated according to a cosine function (1) with 
one year period, $T$, (2)
and a phase, $t_0$, that peaks around $\simeq$ 2$^{nd}$ June (3);
this modulation must only be found
in a well-defined low energy range, where WIMP induced recoils
can be present (4); it must apply to those events in
which just one detector of many actually "fires" 
({\it single-hit} events), since
the WIMP multi-scattering probability is negligible (5);
 the modulation
amplitude in the region of maximal sensitivity is expected 
to be $\lsim$7$\%$  (6). 
For the sake of completeness, we mention that this latter rough limit 
would be larger either
in case the WIMPs would match the
scenario of ref. \cite{Wei01} (because of kinematic effects) or
the scenario of ref. \cite{Fre03} (because of a possible external
contribution to the dark halo from the Sagittarius Dwarf Tidal Stream).
To mimic such a signature 
spurious effects or side reactions
should not only be 
able to account for the whole observed modulation amplitude, 
but also to contemporaneously satisfy all the requirements; 
no one has been found or suggested 
by anyone over about a decade.

As mentioned, to point out the modulation component
of a signal, large mass apparata with suitable performances and
control of the operating conditions are necessary, such as the
$\simeq$ 100 kg highly radiopure NaI(Tl) DAMA/NaI set-up -- which has been
the only experiment able to effectively exploit such a WIMP signature for
 about a decade --
and now the $\simeq$ 250 kg higher radiopure NaI(Tl) DAMA/LIBRA set-up.      
This approach allows to investigate the presence of a 
Dark Matter particle component in the galactic halo
independently on astrophysical, nuclear and particle 
physics modelling. Corollary quests for the
candidate particle require instead to choose a model; therefore, 
the results are not general 
and refer case by case to the considered model framework
\footnote{
We remark that a model framework is identified by the general
astrophysical, nuclear and particle physics assumptions and by the set
of values used for all the experimental and theoretical parameters needed in the calculations
(for example WIMP nature, WIMP couplings, form factors, spin factors, scaling laws, 
quenching factors, halo model, WIMP local velocity, etc.,
which are affected by relevant uncertainties; see for example discussions in ref. \cite{RNC}).}
as it is always the case for results 
given in form of exclusion plots (which, thus, have no generality) and
for the specific parameters of a WIMP candidate (mass and cross sections) derived 
from the indirect searches (for some discussion see \cite{RNC}). 

The presence of a model independent positive evidence in the data of DA\-MA/NaI has been firstly 
reported by the DAMA collaboration at
the TAUP conference in 1997 \cite{taup97} and published also in \cite{Mod1}, confirmed in \cite{Mod2,Ext}, 
further confirmed in \cite{Mod3,Sist,Sisd,Inel,Hep} 
and conclusively  
confirmed, at end of experiment, in 2003 \cite{RNC}.
Corollary model dependent quests for a candidate
particle have been carried out in some of the many possible model
frameworks and have been improved with time. 
In particular, some scenarios either 
for mixed spin-independent (SI) and spin-dependent (SD) 
coupled WIMPs or for purely SI coupled WIMPs or for purely SD 
\footnote{We remind that JHEP 0107 (2001) 044 is not at all in conflict 
with a purely SD solution
since it considered only two particular purely SD couplings (of the many possible) 
in a strongly
model dependent context and using modulation amplitudes valid instead only in a
 particular
purely SI case. Moreover, the mixed SI \& SD case was not involved at all 
in that discussion.} coupled WIMPs 
have been considered in some of the many possible model frameworks
as well as the case of WIMPs with {\em preferred inelastic} scattering.

Finally, some comments on the claims for contradiction 
by Cdms-I, Edel\-weiss-I, Zeplin I can be found in ref. \cite{RNC}\footnote{Here we just remind e.g. 
that no model independent comparison is possible among those experiments and DAMA/NaI 
because of the
different methodological approaches, of the different target nuclei, etc..
As regards possible model dependent comparisons, those experiments give result 
in a single purely SI model framework with fixed/selected assumptions,
neglecting experimental and theoretical
uncertainties and -- as also done in this conference -- have quoted so far the DAMA/NaI 
result in an incorrect , unupdated and incomplete way and they have been and are 
ignoring the existence of other solutions.},
where some recent positive hints --
not in contradiction with the DAMA/NaI result -- from 
Dark Matter indirect searches are also summarised.

\section{Final DAMA/NaI model independent result 
on WIMP annual modulation signature over 7 annual cycles}

A model independent investigation of the WIMP annual modulation
signature has been realised by exploiting the time behaviour of the
experimental residual rates of the {\it single-hit} events in the lowest energy region 
over seven annual cycles (total exposure: 107731 kg $\times$ day) \cite{RNC}, 
as performed in the past on partial exposures. 

The exposure of each annual cycle, DAMA/NaI-1 to -7,
is reported in Table \ref{tab:tab1}, where the exposure of the DAMA/NaI-0 running period 
(partially overlapped with DAMA/NaI-1) is quoted as well.

\begin{table}[!ht]
\vspace{-0.2cm}
\caption{Exposure of each annual cycle (DAMA/NaI-1 to -7) and of 
 the DAMA/NaI-0 running period
\cite{Psd96,Mod1,Mod2,Ext,Mod3,Sist,Sisd,Inel,Hep,RNC}. See text.}
\vspace{-0.1cm}
\begin{center}
\scriptsize
\begin{tabular}{|c|c|}
\hline 
Periods  & Exposure (kg $\times$ day) \\
\hline\hline \hline
DAMA/NaI-0 & 4123.2           \\
\hline\hline \hline
DAMA/NaI-1 & 3363.8 (winter)   \\
           & + 1185.2 (summer) \\
\hline
DAMA/NaI-2 & 14962    \\
 & (Nov. $\rightarrow$ end of July) \\
\hline
DAMA/NaI-3 & 22455            \\
           & (middle Aug. $\rightarrow$ end of Sept.) \\
\hline
DAMA/NaI-4 & 16020           \\
           & (middle Oct. $\rightarrow$ middle Aug.) \\
\hline
DAMA/NaI-5 &    15911                         \\ 
           & (Aug. $\rightarrow$ end of July) \\
\hline
DAMA/NaI-6 &    16608                         \\ 
           & (Nov. $\rightarrow$ end of July) \\
\hline
DAMA/NaI-7 &    17226                         \\ 
           & (Aug. $\rightarrow$ end of July) \\
\hline
{\em Total 1-7} &    107731                         \\
\hline
\end{tabular}
\end{center}
\label{tab:tab1}
\vspace{-0.4cm}
\end{table}

\begin{figure}[ht]
\begin{center}
\vspace{-0.7cm}
\includegraphics[height=10cm]{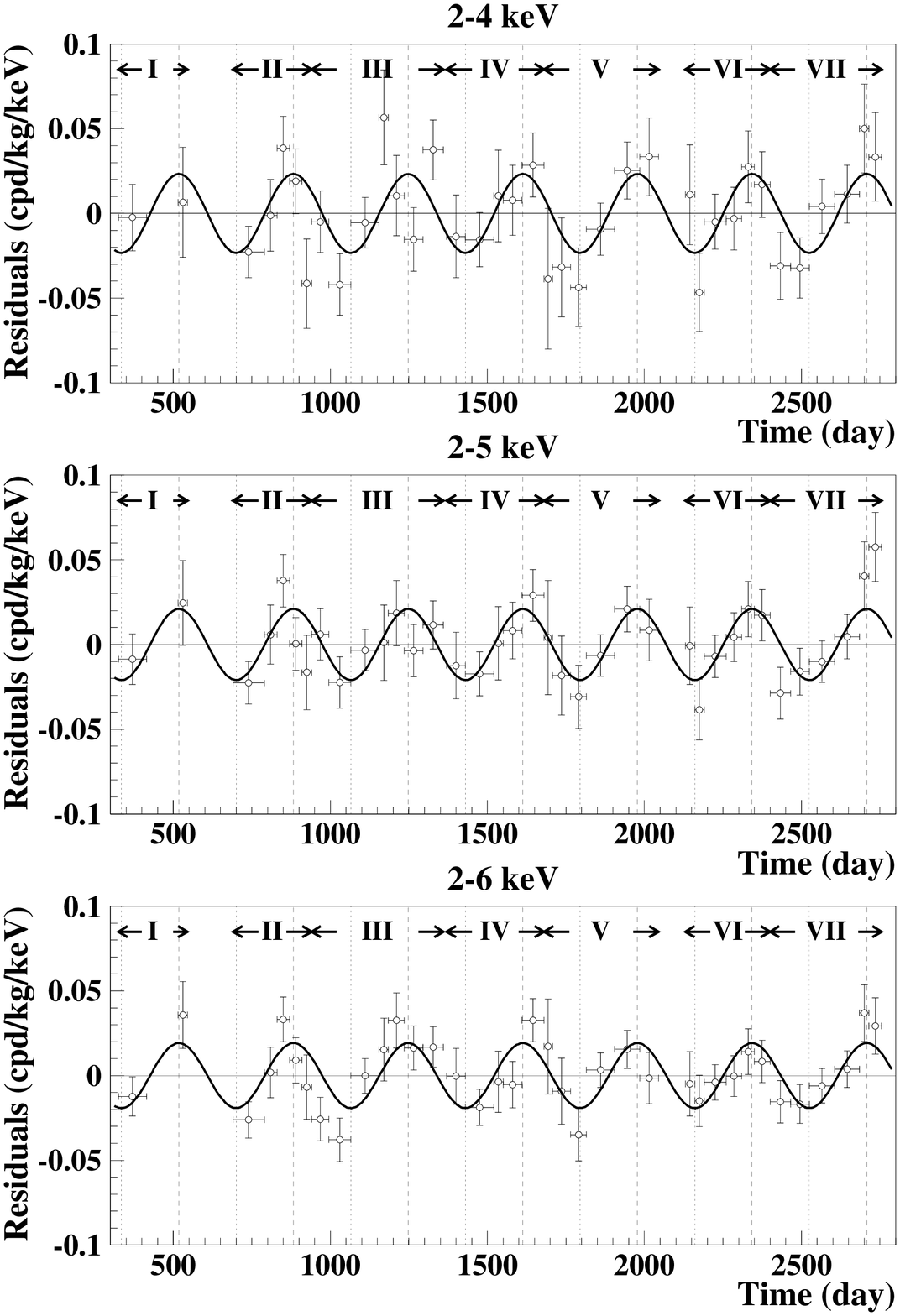}
\includegraphics[height=4.5cm]{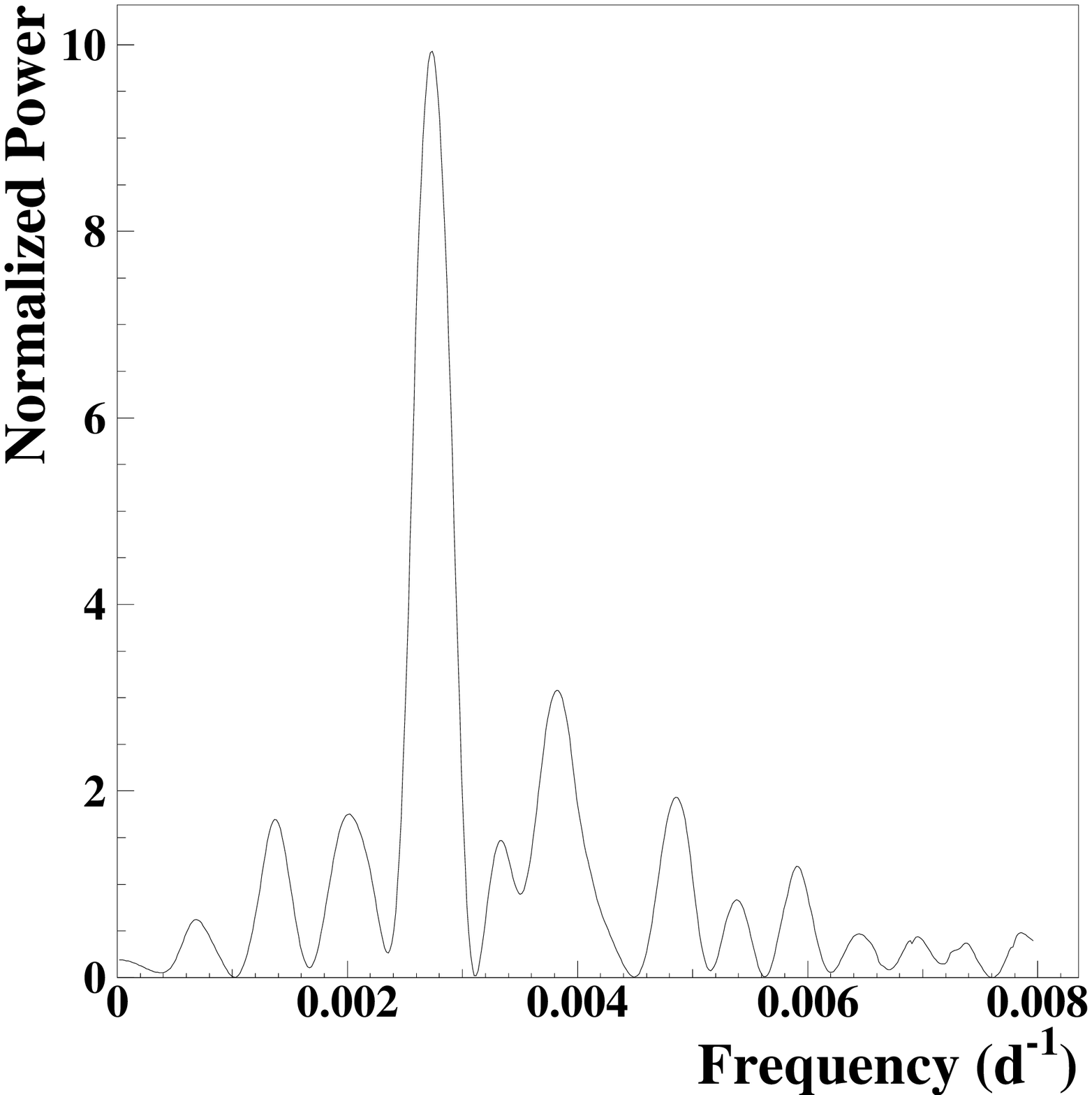}
\end{center}
\vspace{-0.5cm}
\caption{{\it On the left}: experimental 
residual rate for {\it single-hit} events in the (2--4), (2--5) 
and (2--6) keV energy intervals 
as a function of the time over 7 annual cycles (total exposure 
107731 kg $\times$ day); end of data taking July 2002.
The experimental points present the errors as vertical
bars and the associated time bin width as horizontal bars. The 
superimposed curves represent the cosinusoidal functions 
behaviours expected for a WIMP signal 
with a period equal to 1 year and phase at $2^{nd}$ June; the modulation
amplitudes have been obtained by best fit \cite{RNC}. {\it On the right}:
power spectrum of the measured (2--6) keV {\it single-hit} residuals calculated according to ref.
\cite{Lomb}, including also the treatment of the experimental
 errors and of the time binning.
As it can be seen, the principal mode corresponds to a
frequency of $2.737 \cdot 10^{-3}$ d$^{-1}$, 
that is to a period of $\simeq$ 1 year.}
\label{fig1_2}
\vspace{-0.4cm}
\end{figure}
This model independent approach on the data of the seven annual cycles 
offers an immediate evidence of the presence of 
an annual modulation of the rate of the {\it single-hit} events
in the lowest energy region  
as shown in Fig. \ref{fig1_2} -- {\em left}, where the time behaviours 
of the measured (2--4), (2--5) and (2--6) keV
{\it single-hit} events residual rates are
depicted. 

The data favour the presence of a modulated cosine-like behaviour 
($A \cdot$ cos$\omega (t-t_0)$) at 6.3 $\sigma$ C.L. 
and their fit for the (2--6) keV 
larger statistics energy interval offers modulation amplitude equal to
$(0.0200 \pm
0.0032)$ cpd/kg/keV,
$t_0 = (140 \pm 22)$ days and
$T = \frac{2\pi}{\omega} = (1.00 \pm 0.01)$ year, 
all parameters kept free in the fit.
The period and phase agree with those expected in the case of a
WIMP induced effect ($T$ = 1 year and $t_0$ roughly at 
$\simeq$ 152.5$^{th}$ day of the year). 
The $\chi^2$ test on the (2--6) keV residual rate in Fig. \ref{fig1_2} -- {\em left}
disfavours the hypothesis of unmodulated behaviour giving a 
probability of $7 \cdot
10^{-4}$ ($\chi^2/d.o.f.$ = 71/37).
We note that, for simplicity, in Fig. \ref{fig1_2} -- {\em left}  
the same time binning already considered in ref. \cite{Mod3,Sist} has been
used; the result of this approach is similar by choosing other 
time binnings.
The experimental residuals given in Fig. \ref{fig1_2} -- {\em left} have also been fitted, 
according to the previous procedure,
fixing 
the period at 1 year and the phase at $2^{nd}$ June; 
the best fitted modulation
amplitudes
are: $(0.0233 \pm 0.0047)$ cpd/kg/keV for the (2--4) keV 
energy interval, $(0.0210 \pm 0.0038)$
cpd/kg/keV for the (2--5) keV energy interval,
$(0.0192 \pm 0.0031)$ cpd/kg/keV for the (2--6) keV energy interval,
respectively.

The same data have also been investigated by a Fourier analysis
(performed according to ref.
\cite{Lomb} including also the treatment of the experimental 
errors and of the time binning), obtaining
the result shown in Fig.~\ref{fig1_2} -- {\em right}, 
where a clear peak corresponding to a period of 1 year is evident.

\begin{figure}[!ht]
\vspace{-0.8cm}
\centering
\includegraphics[height=4.5cm]{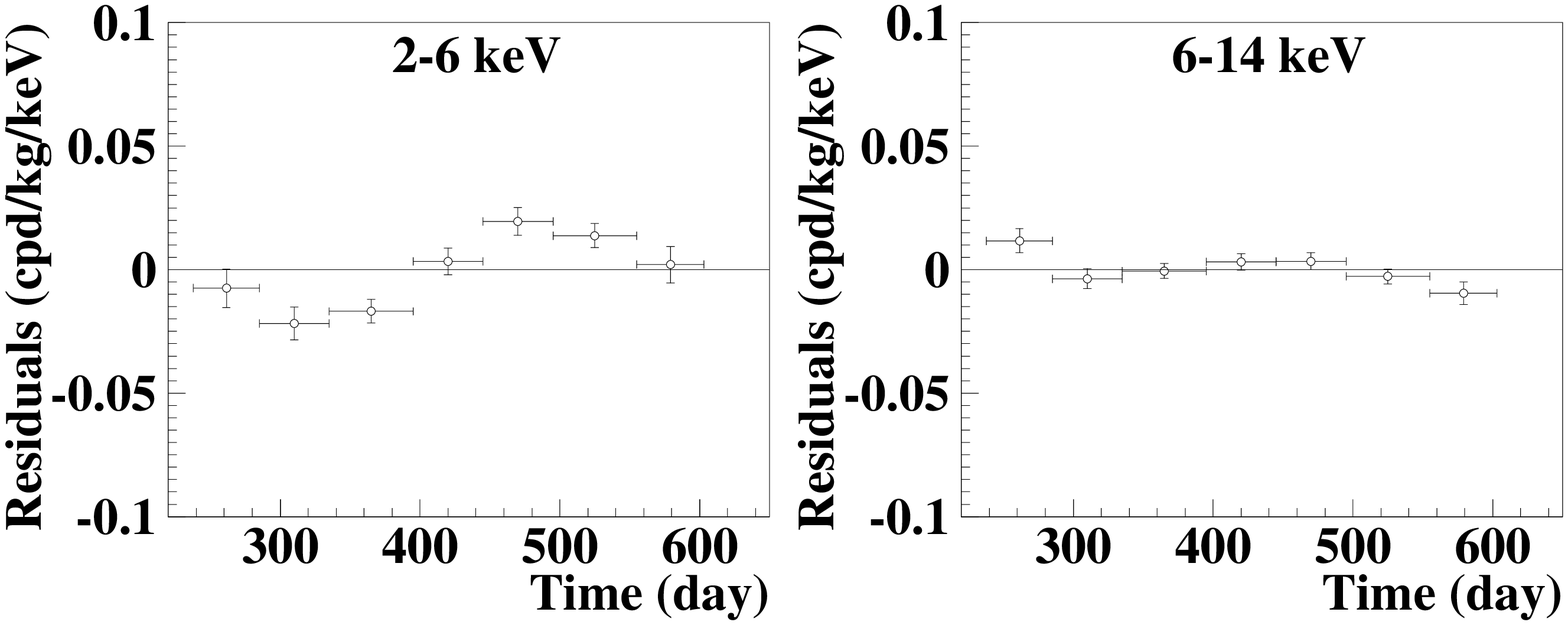}
\vspace{-0.4cm}
\caption{Experimental {\it single-hit}
residual rate from the total exposure of
107731 kg $\times$ day as in a single annual cycle. The
experimental points present the errors as vertical
bars and the associated time bin width as horizontal bars.
The initial time is taken at August 7$^{th}$.
Fitting the data with a cosinusoidal function with period 
of 1 year and phase at 152.5 days,
the following amplitudes are obtained: 
$(0.0195 \pm 0.0031)$ cpd/kg/keV and $-(0.0009 \pm 0.0019)$
cpd/kg/keV, respectively. Thus, a clear modulation 
is present in the lowest energy region, while it
is absent just above.}
\label{fig3}  
\vspace{-0.3cm}
\end{figure}

In Fig. \ref{fig3} the experimental {\it single-hit} residual rate 
from the total exposure of
107731 kg $\times$ day is presented, as in a single 
annual cycle, for two different energy intervals; as it
can be seen the modulation is clearly present in the (2--6) keV 
energy region, while 
it is absent just above.

Finally, a suitable statistical analysis has shown that the 
modulation amplitudes are statistically well distributed 
in all the crystals,
in all the data taking periods and  
considered energy bins \cite{RNC}.

\vspace{0.3cm}
\noindent{\bf On the investigation of possible systematic effects and side reactions.}
As previously mentioned, to mimic the annual modulation
signature a systematic effect or side reaction should not only be 
able to account for the whole observed modulation amplitude, 
but also able to satisfy the requirements of a WIMP induced effect.
A careful investigation of all the known possible sources
of systematics and side
reactions has been regularly carried out and published at time of each data 
release.
In particular, detailed quantitative discussions can be found 
in ref. \cite{Sist,RNC} and will not be repeated here \footnote{
We take this opportunity only to comment that the
sizeable discussions reported e.g. in \cite{Sist,RNC}
already demonstrated that
a possible modulation of neutron flux
(possibly observed by the ICARUS coll.
as reported in the ICARUS internal report TM03-01)
cannot quantitatively
contribute to the DAMA/NaI observed modulation amplitude,
even if the neutron flux would
be assumed to be 100 times larger than measured at LNGS by several authors
with different techniques over more than 15 years; in addition, as widely known, 
it cannot satisfy all the peculiarities of the signature.}.
As it can be seen there, no systematic effect or side reaction able to mimic a WIMP induced effect
has been found. 

\vspace{0.3cm}
\noindent{\bf A further result: the {\it multiple-hits} events.}
As a further relevant investigation, 
the {\it multiple-hits} events also collected during the DAMA/NaI-6 
and 7 running periods
(when each detector was equipped with
its own Transient Digitizer with a dedicated renewed electronics)
have been studied and 
analysed by using the same identical hardware and the same 
identical software 
procedures as for the case of the {\it single-hit} events.
The {\it multiple-hits} events class -- on the contrary of the {\it single-hit} one
-- does not include events induced by
WIMPs since the probability that a WIMP scatters 
off more than one detector is
negligible.  

\begin{figure}[ht]
\centering
\vspace{-1.cm}
\includegraphics[height=5.cm]{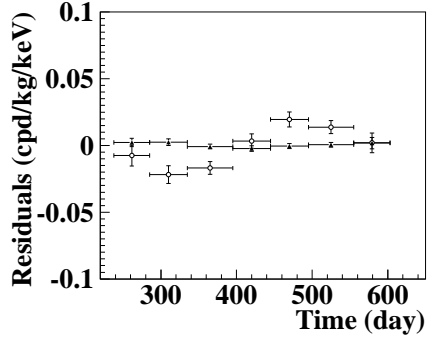}
\caption{Experimental residual rates over seven annual cycles
for {\it single-hit} events (open circles) -- class of events 
to which WIMP events belong --
and over the last two annual cycles for {\it multiple-hits} 
events (filled triangles)
-- class of events to which WIMP events do not belong -- in the
(2--6) keV cumulative energy interval. They have been obtained 
by considering for each class of events the 
the data as collected 
in a single annual cycle and using in both cases the same 
identical hardware and the same identical software
procedures. The initial time is taken on August 7$^{th}$.}
\label{fg:fig6}
\vspace{-0.4cm}
\end{figure}

Fig. \ref{fg:fig6} shows the behaviour of the  
residual rate of {\it multiple-hits} events in the (2--6) keV 
energy interval measured during 
the DAMA/NaI-6 and -7 running periods
as a function of the time in a year.
It is compared with the residual rate of the {\it single-hit} events 
measured in the same energy interval with the total exposure.
Fitting these data with the function $A \cdot$ cos$\omega (t-t_0)$
with period of 1 year and phase at 152.5 days, the following
amplitudes are obtained: $A=(0.0195\pm0.0031)$ cpd/kg/keV 
and $A=-(3.9\pm7.9) \cdot 10^{-4}$ cpd/kg/keV for {\it single-hit} 
and {\it multiple-hits} residual rates, respectively.
Thus, a 6.3 $\sigma$ C.L. evidence of annual modulation is present in 
the {\it single-hit} residuals (events class to which the WIMP-induced recoils
belong), while it is absent in the 
{\it multiple-hits} residual rate (event class to which only background events belong).
Since the same identical hardware and the same identical software procedures have been 
used to analyse the two classes of events,
the obtained result offers an additional strong support for the presence of 
Dark Matter particles in the galactic halo further excluding any side effect 
either from hardware or from software procedures or from background.

\vspace{0.3cm}
\noindent {\bf Conclusion.}
In conclusion, the presence of an annual modulation in the 
{\it single-hit} events residual rate 
in the lowest energy interval (2 -- 6) keV 
satisfying all the features expected for a WIMP component 
in the galactic halo
is supported by the data of the seven annual cycles at 6.3 $\sigma$ C.L..
This is the experimental result of DAMA/NaI. It is model 
independent; no other experiment whose result can be 
directly compared with this one is available so far in the field of Dark Matter
investigation.

\section{Corollary result: quests for a candidate particle in some model
frameworks with the data of the seven annual cycles}

On the basis of the obtained model independent result, corollary 
investigations can also be
pursued on the nature and coupling of the WIMP candidate.
This latter investigation is instead model
dependent and -- considering the large uncertainties which exist on the 
astrophysical, nuclear and particle physics
assumptions and on the parameters needed in the calculations -- has no general
meaning (as it is also the case -- as mentioned above -- of exclusion plots and 
of the WIMP parameters evaluated in indirect
search experiments). Thus, it should be
handled in the most general way as we have preliminarily
 pointed out with time in the past
\cite{Mod1,Mod2,Ext,Mod3,Sist,Sisd,Inel,Hep} and we have discussed
in some specific details in ref. \cite{RNC};
other efforts on this topic are also in progress.
Candidates, kinds of WIMP couplings with ordinary matter and implications, 
cross sections, nuclear form factors, spin factors, 
scaling laws, halo models, priors, etc. are discussed in details in ref. \cite{RNC}
and we invite the reader to this reference since these arguments are necessary
to correctly understand the results obtained in corollary quests
and the real validity of any claimed 
model dependent comparison in the field.
The results presented in ref. \cite{RNC} and summarised here
are, of course, not exhaustive of the many possible scenarios
which at present level of knowledge 
cannot be disentangled. Some of the open questions are: i) which is 
the right nature for the WIMP particle \footnote{
Several candidates fulfil the cosmological and particle Physics
requirements necessary in order to be considered 
as a Dark Matter candidate particle: not only the neutralino foreseen in the 
supersymmetric theories, 
but also a heavy neutrino of a 4$^{th}$ family 
(there is still room for it as reported in literature), the sneutrino 
in the scenario described by \cite{Wei01} 
(providing -- through the transition from lower to upper mass 
eigenstate -- {\em preferred inelastic} scattering with target-nuclei), 
the ``mirror'' Dark Matter \cite{foot}, etc..
Moreover, in principle whatever Weakly Interacting, neutral, (quasi-)stable 
and Massive (whose acronym is WIMP) particle, even not yet foreseen 
by a theory, can be a suitable candidate.
As regards in particular the neutralino, we note that 
the theories have not stringent predictive capability for its 
cross sections and for its mass because of the large number of free
parameters in the theory and of the several assumptions required; thus,
e.g. the expectations for its nuclear cross sections span 
over several orders of magnitude as it can also be seen in literature.
In addition, we take this occasion to remind that the neutralino
has both SI and SD couplings with the ordinary matter.};
ii) which is its right couplings with ordinary matter (mixed SI\&SD, 
purely SI, purely SD or 
{\em preferred inelastic})
iii) which are the right form factors and
related parameters for each target nucleus; iv) 
which is the right spin factor for each target nucleus 
(some nuclei are disfavoured to some kinds of interactions; for example,
in case of an interaction with SD component 
even a nucleus sensitive in principle
to SD interaction could be blinded by the spin factor 
if unfavoured by the $\theta$ value\footnote{We remind that 
$tg \theta = a_n/a_p$ is the ratio between the WIMP-neutron and the 
WIMP-proton effective SD coupling strengths, $a_n$ and $a_p$,
 respectively \cite{Sisd,RNC};
$\theta$ is defined in the [0,$\pi$) interval.});
v) which are the right scaling laws (let us consider as an 
example that even in a MSSM framework with purely SI interaction the scenario could be
drastically modified as discussed recently in ref. \cite{Kam03}); 
vi) which is the right halo model and
related parameters; vii) which are the right values 
of the experimental parameters within their uncertainties; etc.
As an example, we remind that not only large differences 
in the measured rate can be expected when using
target nuclei sensitive to the SD component of the interaction (such as e.g.
$^{23}Na$ and $^{127}I$) with respect to those largely insensitive to such a
coupling 
(such as e.g. $^{nat}Ge$ and $^{nat}Si$), but also when using different 
target nuclei although all -- in principle --
sensitive to such a coupling (compare e.g. the Xenon and Tellurium cases 
with the Sodium and Iodine cases) \cite{RNC}. 

In the following some of the results discussed
for some of the many possible model dependent 
quests for a WIMP candidate are briefly reminded \cite{RNC}.
In particular, they have been obtained from the data
collected during all the seven annual cycles,  
considering the halo models summarized in \cite{Hep,RNC} 
for three of the possible values 
of the local velocity $v_0$: 170 km/s, 220 km/s and 270 km/s.
The escape velocity has been maintained at the fixed value: 650 km/s. 
It is worth to note that the
present existing uncertainties on the knowledge of the escape velocity can play a 
relevant role in evaluating allowed regions (and corresponding 
best fit values for WIMP mass and cross section)
e.g. in the cases of {\em
preferred inelastic} WIMPs and of light mass 
WIMP candidates; its effect would be instead marginal at large WIMP masses.
The possible scenarios have been
exploited for those halo models in 
some discrete cases including some of the uncertainties 
which exist in the parameters of the used nuclear form factors and 
in the quenching factors; for the details see ref. \cite{RNC}.
The results summarised here are not exhaustive of the many scenarios
possible at present level of knowledge: e.g. for some other recent ideas 
see the already quoted \cite{Fre03,Kam03}.

For simplicity, here the results of these corollary quests for a
candidate particle
are presented in terms of allowed regions
obtained as superposition of the configurations corresponding
to likelihood function values {\it distant} more than $4\sigma$ from
the null hypothesis (absence of modulation) in each of the several 
(but still a limited number) of the possible 
model frameworks considered here. 
Obviously, larger  
sensitivities than those reported in the following figures would be reached when including
the effect of other existing
uncertainties on the astrophysical, nuclear and particle Physics assumptions and related parameters;
similarly, the set of the best fit values would also be enlarged as well.

As well known, DAMA/NaI is
intrinsically sensitive both to low and high WIMP mass having both a light
(the $^{23}$Na) and a heavy (the $^{127}$I) target-nucleus;
in previous corollary quests
WIMP masses above 30 GeV (25 GeV in ref.
\cite{Mod1}) have been presented \cite{Mod2,Mod3,Sisd,Inel,Hep} 
for few (of the many possible) model frameworks.
However, that bound holds only for neutralino when 
supersymmetric schemes based on GUT assumptions
are adopted to analyse the LEP data \cite{Dpp0}.
Thus, since other candidates are possible and also other scenarios can be considered for the 
neutralino itself as recently pointed out
\footnote{In fact, when the assumption on the gaugino-mass
unification at GUT scale is released neutralino masses down to $\simeq$ 6 GeV 
are allowed \cite{Bo03,lowm,bbpr}.}, the present model dependent lower bound quoted by LEP 
for the neutralino in the
supersymmetric schemes based on GUT assumptions (37 GeV \cite{Dpp}) is simply marked
in the following figures. 
It is worth to note that when this mass limit is adopted, it selects the
WIMP-Iodine elastic scattering as dominant because of the used
scaling laws and of kinematical arguments.
Finally, the prior from DAMA/NaI-0 has properly been considered as well. 

\vspace{0.3cm}
\noindent {\bf WIMPs with mixed SI\&SD interaction.}
The most general scenario of WIMP nucleus elastic interaction, to which the DAMA/NaI target 
nuclei are fully sensitive,
is the one  where both the SI and the SD components of the cross section 
\begin{figure}[ht]
\begin{center}
\vspace{-0.4cm}
\includegraphics[height=9cm]{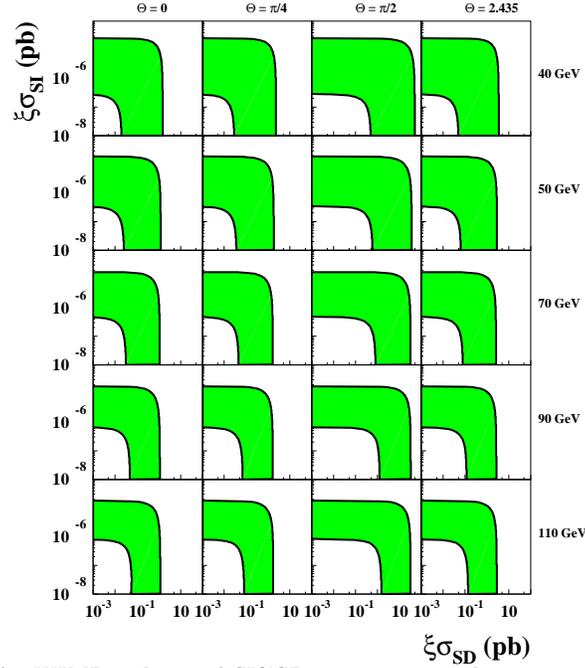}
\end{center}  
\vspace{-0.6cm}
\caption{{\it Case of a WIMP with mixed SI\&SD interaction in the given model
frameworks}. Coloured areas: example of slices (of
the allowed volume) in the plane 
$\xi \sigma_{SI}$ vs
$\xi \sigma_{SD}$ for some of the possible $m_W$ and $\theta$ values.
Inclusion of other existing uncertainties on parameters and models 
would further extend the regions; for example,
the use of more favourable form factors and/or of more favourable
spin factors than the considered ones 
would move them towards lower cross sections.
For details see \cite{RNC}.}
\label{fg:pan_sisd}
\vspace{-0.4cm}
\end{figure}
are present. Thus, as first we introduce here the case for a
candidate with both SI and SD couplings to ordinary matter \cite{RNC} 
similarly as we did
in the past also in ref. \cite{Sisd} on partial exposure.
In this general scenario the space of the free parameters is 
a 4-dimensional volume defined by $m_W$, $\xi \sigma_{SI}$ 
\footnote{$\xi$ ($\xi\leq 1$) is defined here as the fractional amount of local WIMP density.},
$\xi \sigma_{SD}$ and $\theta$ (which varies from 0 to $\pi$).
Thus, the general solution would be a four dimensional allowed volume
for each considered model framework.
Since the graphic representation of this allowed volume is quite difficult, 
we only show in Fig. \ref{fg:pan_sisd} the obtained regions in the plane 
$\xi \sigma_{SI}$ vs $\xi \sigma_{SD}$ for some of the possible 
$\theta$ and $m_W$ values in the model frameworks considered here.
In particular, we report just four couplings, which correspond to the
following values of the mixing angle $\theta$: i)  $\theta$ = 0 ($a_n$
=0 and $a_p \ne$ 0 or  $|a_p| >> |a_n|$) corresponding to a particle
with null SD coupling to neutron; ii) $\theta = \pi/4$ ($a_p = a_n$)
corresponding to a particle with the same SD coupling to neutron and
proton; iii)  $\theta$ = $\pi/2$ ($a_n \ne$ 0 and $a_p$ = 0 
or  $|a_n| >> |a_p|$) corresponding to a particle with null SD
couplings to proton; iv) $\theta$ = 2.435 rad ($ \frac {a_n} {a_p}$
= -0.85) corresponding to a particle with SD coupling through $Z_0$
exchange. The case $a_p = - a_n$ is nearly similar to the case iv).  

From the given figures it is clear that at present either a purely SI or a purely 
SD or a mixed SI\&SD configurations
are supported by the experimental data of the seven annual cycles. 

\vspace{0.3cm}
\noindent {\bf WIMPs with dominant SI interaction.}
Generally, the case of purely SI
coupled WIMP is mainly considered in literature. In fact, often the 
spin-independent interaction
with ordinary matter is assumed to be dominant since e.g.
most of the used target-nuclei are practically
not sensitive to SD interactions (as on the contrary
$^{23}$Na and $^{127}$I are) and the theoretical calculations 
are even much more complex and uncertain.

Thus, following an analogous procedure as for the previous case, we have exploited for the same model
frameworks the purely SI scenario. 
In this case the free parameters are two: $m_W$ and $\xi \sigma_{SI}$. 

\begin{figure}[!ht]
\vspace{-0.6cm}
\begin{center}
\includegraphics[height=4.5cm]{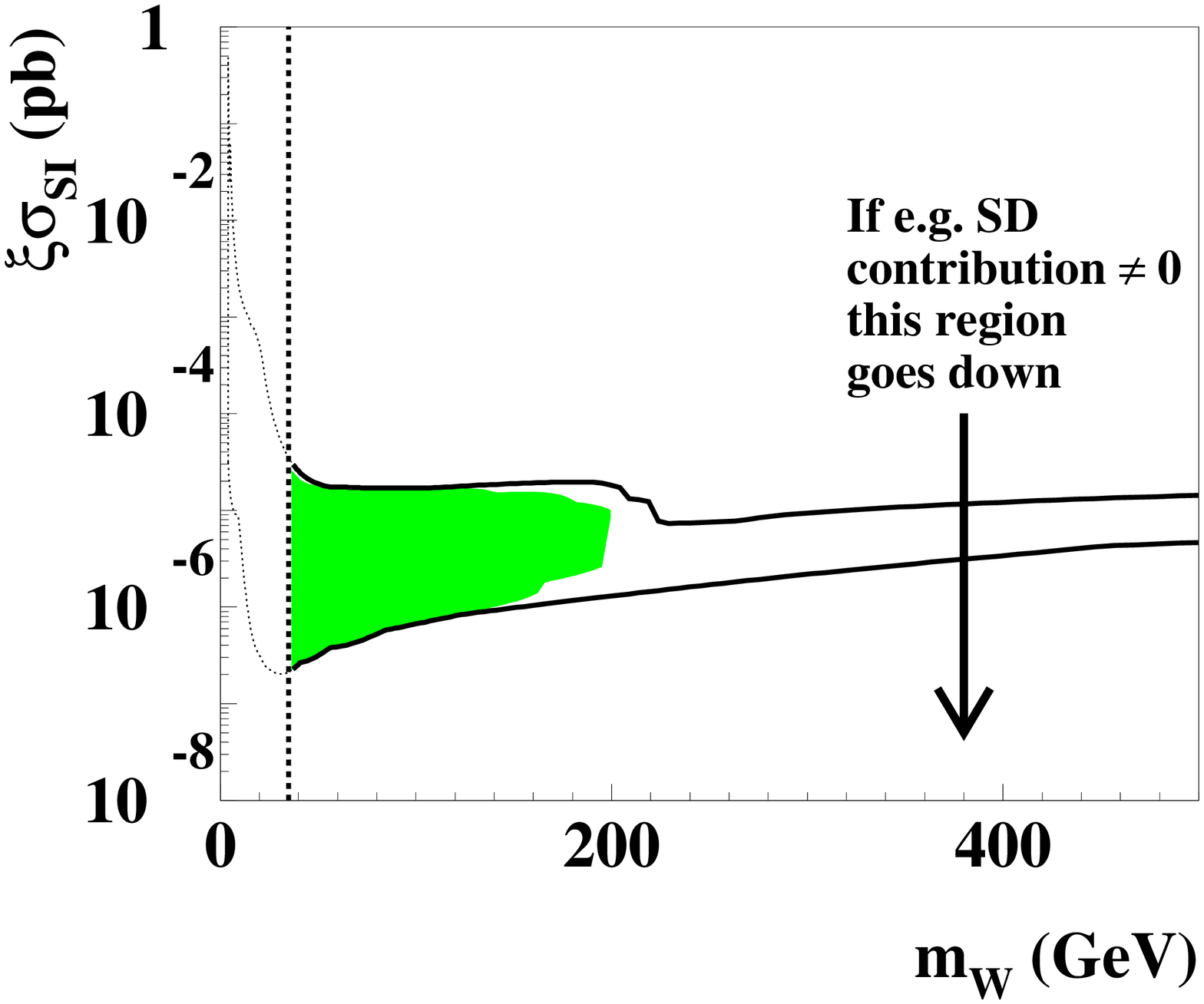}
\includegraphics[height=4.5cm]{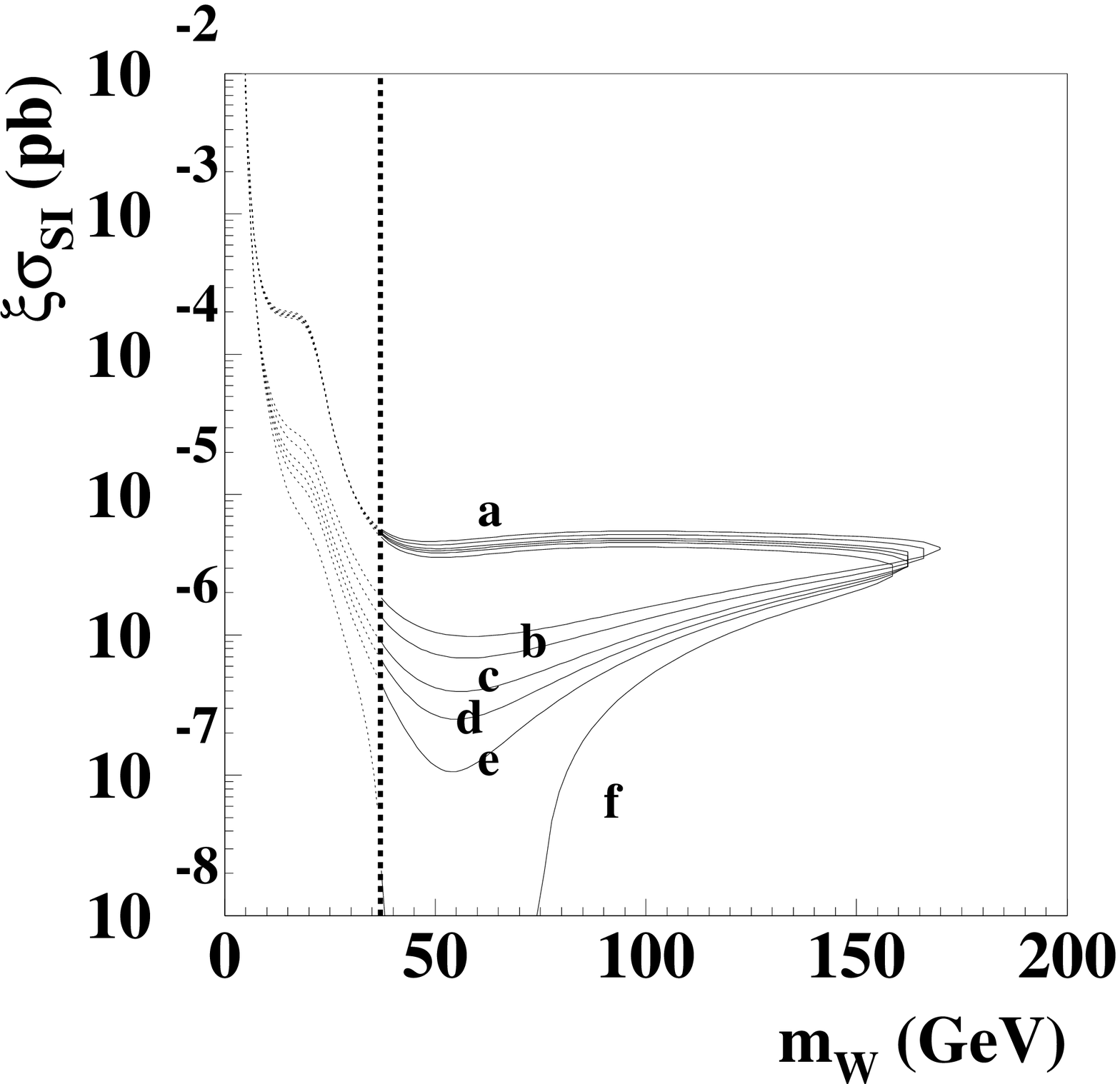}
\end{center}  
\vspace{-0.6cm}
\caption{{\it On the left : Case of a WIMP with dominant SI interaction for the given model
frameworks.} Region allowed in the plane ($m_W$, $\xi
\sigma_{SI}$). The vertical dotted line represents a bound in case
of a neutralino candidate when 
supersymmetric schemes based on GUT assumptions
are adopted to analyse the LEP data; 
the low mass region is allowed for neutralino when other schemes are 
considered and for every other WIMP candidate; see text.
While
the area at WIMP masses above
200 GeV is allowed only for few configurations, the lower one is allowed by most configurations 
(the colored region gathers only those above 
the vertical line) \cite{RNC}.
The inclusion of other existing uncertainties on parameters and models 
would further extend the region; for example, 
the use of more favourable SI form factor for Iodine
alone would move it towards lower cross sections.
{\it On the right :
Example of the effect induced by the inclusion of a SD component
different from zero on  
allowed regions given in the plane $\xi\sigma_{SI}$ vs $m_W$.}
In this example the
Evans' logarithmic axisymmetric $C2$ halo model with
$v_0 = 170$ km/s, $\rho_0$ equal to the maximum value for this model 
and a given set of the parameters' values (see \cite{RNC}) have been considered.
The different regions refer to different SD contributions for the particular case of 
$\theta = 0$:
$\sigma_{SD}=$ 0 pb (a), 0.02 pb (b), 0.04 pb (c), 0.05 pb (d),
0.06 pb (e), 0.08 pb (f). Analogous situation is found for the other 
model frameworks.}
\label{fg:fig_si}
\vspace{-0.4cm}
\end{figure}

In Fig. \ref{fg:fig_si} -- {\em left} the region allowed in the plane $m_W$ and $\xi
\sigma_{SI}$ for the considered model frameworks is reported. 
The configurations below the vertical line are of interest 
for neutralino when the assumption on the gaugino-mass
unification at GUT scale is released
and for every other kind of WIMP candidate.
As shown in Fig. \ref{fg:fig_si} -- {\em left}, also WIMP masses above 
200 GeV are allowed for some configurations; details can be found in
ref. \cite{RNC}.
Of course, best fit values of cross section and WIMP mass 
span over a large range in the considered frameworks.

Let us now point out, in addition, that 
configurations with $\xi \sigma_{SI}$ even much lower than those shown in
Fig. \ref{fg:fig_si} -- {\em left} are accessible in case an even small SD contribution 
is present in the interaction. 
This possibility is clearly pointed out
in Fig. \ref{fg:fig_si} -- {\em right} where an example of allowed regions in the plane
($m_W$, $\xi \sigma_{SI}$) corresponding to different SD contributions
is reported for the case $\theta =0$.
As it can be seen,
increasing the SD contribution the allowed regions 
involve SI cross sections much lower than 
$10^{-6}$ pb. It can be noted that for $ \sigma_{SD} \geq 0.08$
pb the annual modulation effect observed is also compatible 
-- for $m_W \simeq 40-75$ GeV -- with a WIMP
candidate with no SI interaction at all.
Analogous situation is found for the other 
model frameworks.

\vspace{0.3cm}
\noindent{\bf WIMPs with dominant SD interaction.}
Let us now focus on the case of a candidate with purely SD 
\begin{figure}[!ht]
\begin{center}
\vspace{-0.6cm}
\includegraphics[height=4.cm]{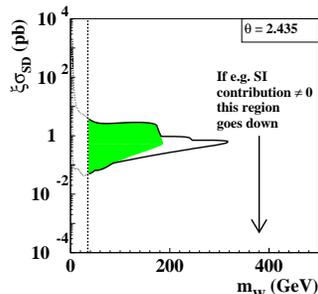}
\end{center}  
\vspace{-0.6cm}
\caption{{\it Case of a WIMP with dominant SD interaction in the given model
frameworks.} An example of the
region allowed in the plane ($m_W$, $\xi \sigma_{SD}$); here 
$\theta = 2.435$, $Z_0$ coupling, ($\theta$ is defined in the $[0,\pi)$ range). 
For the definition of the vertical line and of the coloured
area see previous figure caption; see also text.
Inclusion of other existing uncertainties on
parameters and models (as discussed in ref. \cite{RNC}) 
would further extend the SD allowed regions. For example, 
the use of more favourable SD form factors 
and/or more favourable spin factors
would move them towards lower cross sections.
Values of $\xi \sigma_{SD}$ lower than those corresponding to these allowed
regions are possible also e.g. in case of an even small
SI contribution (see ref. \cite{RNC}).}
\label{fg:fig_puresd}
\vspace{-0.4cm}
\end{figure}
coupling to which DAMA/NaI is 
-- as mentioned --  fully sensitive.

When the SD component is different from zero, 
a very large number of possible configurations is available.
In fact, in this scenario the space of free parameters 
is a 3-dimensional volume defined by $m_W$, $\xi \sigma_{SD}$ and 
$\theta$ (which can vary from 0 to $\pi$).
Here, just as an example we show the results obtained 
only for a particular coupling, which correspond to 
a mixing angle $\theta$ = 2.435 ($Z_0$ coupling); see 
Fig. \ref{fg:fig_puresd};
other configurations are possible varying the $\theta$ value.
The area at WIMP masses above
200 GeV is allowed for low local velocity  and all considered
sets of parameters
by the Evans' logarithmic $C2$
co-rotating halo model \cite{RNC}.  
Moreover, the accounting for
the uncertainties e.g. on the form factors and/or on the spin factors as well as
different possible formulations of the SD form factors would extend 
the allowed regions, e.g. towards
lower
$\xi\sigma_{SD}$ values.
Finally, $\xi \sigma_{SD}$ lower than those corresponding to the
regions shown in Fig. \ref{fg:fig_puresd} are possible also e.g. in case of an even small
SI contribution (see ref. \cite{RNC}).

\vspace{0.3cm}
\noindent{\bf WIMPs with {\em preferred inelastic} interaction.}
An analysis considering the same model frameworks has been carried out 
for the case of WIMPs with {\em preferred inelastic} interaction \cite{Wei01} 
as we did also in the past in ref. \cite{Inel} on partial exposure.

\begin{figure}[!htb]
\begin{center}
\vspace{-0.3cm}
\includegraphics[height=7cm]{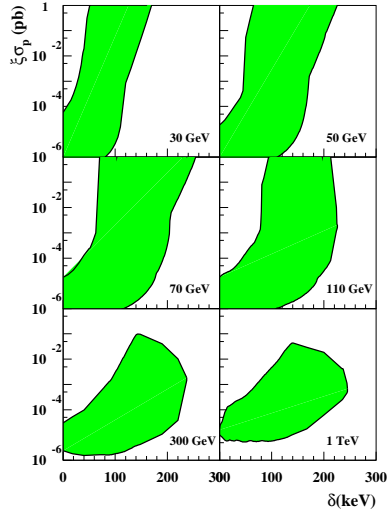}
\end{center}  
\vspace{-0.6cm}
\caption{{\it Case of a WIMP with {\em preferred inelastic} interaction
in the given model frameworks.} Examples of
slices (coloured areas) of the allowed volumes
($\xi \sigma_p$, $\delta$, $m_W$) for some $m_W$ values
for WIMP with {\em preferred inelastic} interaction. 
Inclusion of other existing uncertainties 
on parameters and models 
would further extend the regions; for example,
the use of a more favourable SI form factor for Iodine 
and different escape velocity would move them towards lower cross sections
\cite{RNC}.}
\label{fg:fig_inel}
\vspace{-0.4cm}
\end{figure}

In this inelastic Dark Matter scenario an allowed volume 
in the space ($\xi \sigma_p$, $m_W$, $\delta$) 
is obtained; $\delta$ is the mass splitting of the WIMP particle 
which can be excited following an inelastic interaction
\cite{Wei01,Inel,RNC}. 
 For simplicity, Fig. \ref{fg:fig_inel} shows slices of such an allowed 
volume at some given WIMP masses. 
There the superpositions
of the allowed regions obtained, when varying the model framework
within the considered set \cite{RNC}, are shown for each $m_W$. 
We remind that in these calculations
$v_{esc}$
has been assumed at fixed value
(as in the previous cases), while its present uncertainties can play a significant 
role in this scenario of WIMP with {\em preferred inelastic} scattering.

\vspace{0.3cm}
\noindent{\bf Conclusion.}
In this section we have briefly summarized some quests for the candidate particle
in some of the many possible scenarios. 
We further stress that, although several scenarios have been investigated,
these corollary analyses are not exhaustive at all because of the present poor knowledge
on many astrophysical, nuclear and particle Physics needed assumptions;
moreover, additional scenarios can also be possible as 
also shown e.g. by some recent papers appeared in literature.
Other model dependent quests are already under consideration.

\section{The second generation: DAMA/LIBRA}

In 1996 a ton set-up was proposed by Bernabei et al.; as a consequence
a new R\&D for NaI(Tl) radiopurification has been carried out
and the second generation set-up DAMA/LIBRA ($\simeq$250 kg NaI(Tl)) 
has been funded and realised as an intermediate step.
This R\&D with 
Crismatec-St. Gobain company has exploited for the first time 
new chemical/physical radiopurification procedures in NaI 
and TlI powders already selected for radiopurity. 
In addition, new selected materials and set-up components
as well as new protocols have been employed for building DAMA/LIBRA.

In 2002 ended the production of detectors and of  new parts of the installation
for DAMA/LIBRA. 
Thus, after July 2002 -- at the completion of its data taking -- DAMA/NaI 
was fully dismounted and the installation of the new DAMA/LIBRA started.
The experimental site as well as many components of the installation itself 
were implemented
(environment, shield of PMTs, wiring, HP Nitrogen system, cooling water of air        
conditioner, electronics and DAQ, etc...).
Before the installation, all the Cu parts were 
chemically etched following a devoted protocol and maintained in HP
Nitrogen atmosphere.
In addition, all the procedures performed during the dismounting of DAMA/NaI 
and the installation of DAMA/LIBRA were carried out in HP Nitrogen atmosphere.
This was realised by using a Scuba system (a self-contained underwater breathing apparatus) 
modified in order to avoid that the entire breath is
expelled into the surrounding air when the operator exhales; the air cylinders were 
kept five meters away and the output 
line was two meters long.

\begin{figure}[!htpb]
\centering
\vskip -0.8cm
\includegraphics[width=180pt]{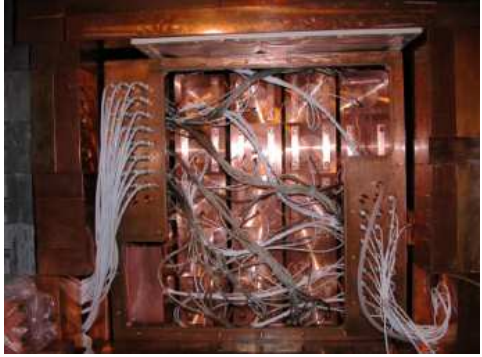}
\vskip -0.1cm
\caption{DAMA/LIBRA: a view of the detectors inside the inner Cu box at end of installation (photo
taken in HP Nitrogen atmosphere).
All the used materials have been deeply selected for radiopurity.}
\vskip -0.5cm
\label{fg:mezzaluna}
\end{figure}

The new DAMA/LIBRA, having an exposed mass of $\simeq 250$
kg, an higher overall radiopurity and improved performances, 
offers an increased experimental
sensitivity to further investigate the DAMA/NaI observed effect and 
to improve investigations on the nature of the candidate particle
trying to disentangle among different possible astrophysical, nuclear 
and particle Physics models as well as other new possible scenarios.
See just as an example of new ideas the case of the {\it mirror} Dark Matter
\cite{foot}, the case of 
a contribution to the dark halo from the Sagittarius Dwarf Tidal Stream \cite{Fre03},
the case of possible different nuclear scaling laws even for the neutralino candidate
in MSSM \cite{Kam03}, etc.. 
Moreover, the low background DAMA/LIBRA offers a powerful tool for the Dark Matter 
investigation in the future since it is e.g.
sensitive: i) both to low (through interaction on $^{23}$Na) 
and to high (through interaction on $^{127}$I) mass Dark Matter particles; ii) both
to mixed SI\&SD, to purely SI, to purely SD couplings and to {\em preferred inelastic}
scattering as well as to other possible kind of Dark Matter candidates 
(e.g. {\it mirror} Dark Matter).

DAMA/LIBRA has started the preliminary data taking on March 2003
and it has been planned to run for several years.
At present a new R\&D effort toward the possible NaI(Tl) ton set-up
has been funded and related works have already been started.

\section{Conclusion}

DAMA/NaI has been a pioneer experiment running 
at LNGS for about a decade and
investigating as first the WIMP annual modulation signature 
with suitable exposed mass, sensitivity and control of the running parameters. 
During seven independent 
experiments of one year each one, it has pointed out the presence of a modulation
satisfying the many peculiarities of a WIMP induced effect, reaching an evidence at
6.3 $\sigma$ C.L.. No other experiment has so far been in position to give a result
directly comparable in a model independent way with that of DAMA/NaI.
As a corollary, it has also pointed out the complexity of the quest for a WIMP
candidate because of the present poor knowledge on the many astrophysical,
nuclear and particle physics aspects.

After the completion, on July 2002, of the DAMA/NaI data taking,
the second generation DAMA/LIBRA set-up ($\simeq$ 250 kg mass) has been 
installed and preliminarily put in operation, as a result of  
continuous efforts by the DAMA collaboration toward the creation of ultimate radiopure 
NaI(Tl).
DAMA/LIBRA, having a larger exposed mass 
and an higher overall radiopurity, will  
significantly contribute in the incoming years to the
further understanding of the field. 
Presently, new R\&D efforts toward a possible NaI(Tl) ton set-up
have been funded and related works have already been started.

\end{document}